# The ADAPT Tool:
# From AADL Architectural Models to Stochastic Petri Nets through Model Transformation

Ana-Elena Rugina, Karama Kanoun and Mohamed Kaâniche
*LAAS-CNRS, University of Toulouse*
*7 avenue Colonel Roche*
*31077 Toulouse Cedex 4, France*
*Phone: +33(0)5 61 33 62 00, Fax: +33(0)5 61 33 64 11*
*e-mail: ana-elena.rugina@astrium.eads.net; {kanoun, kaaniche}@laas.fr*

## Abstract

*ADAPT is a tool that aims at easing the task of evaluating dependability measures in the context of modern model driven engineering processes based on AADL (Architecture Analysis and Design Language). Hence, its input is an AADL architectural model annotated with dependability-related information. Its output is a dependability evaluation model in the form of a Generalized Stochastic Petri Net (GSPN). The latter can be processed by existing dependability evaluation tools, to compute quantitative measures such as reliability, availability, etc.. ADAPT interfaces OSATE (the Open Source AADL Tool Environment) on the AADL side and SURF-2, on the dependability evaluation side. In addition, ADAPT provides the GSPN in XML/XMI format, which represents a gateway to other dependability evaluation tools, as the processing techniques for XML files allow it to be easily converted to a tool-specific GSPN.*

## 1. Introduction

The increasing complexity of new-generation systems raises major concerns in various critical application domains, in particular with respect to the validation and analysis of performance, timing and dependability-related requirements. Model-driven engineering (MDE) approaches aimed at mastering this complexity during the development process have emerged and are being increasingly used in industry. They address the problem of complexity by promoting reuse and partial or total automation of certain phases of the development process. These engineering approaches must be supported by languages and tools that provide means to ensure that the implemented system complies with its specifications. In particular, the automatic derivation of models allowing the analysis of quality attributes[1] (such as dependability and performance) from modeling languages used in MDE is of primary interest.

The AADL (Architecture Analysis and Design Language) is considered in a number of projects aiming at defining and implementing tool support for MDE. It has received a growing interest from the embedded safety-critical industry (e.g., Honeywell, Rockwell Collins, Lockheed Martin, the European Space Agency, Astrium, Airbus) and has been standardized in 2004 under the auspices of the International Society of Automotive Engineers [1]. AADL provides a standardized textual and graphical notation for describing software and hardware system architectures and their functional interfaces. The serious consideration of AADL by the embedded safety-critical industry is justified by the AADL's advanced support for modeling reconfigurable architectures and for analyzing quality attributes [2].

Several tools have been implemented so far in order to support various analyses based on AADL models. [3] reports the implementation of a tool that automatically translates an AADL model into the real time process algebra ACSR (Algebra of Communicating Shared Resources), for schedulability analysis. Schedulability and memory requirements can be analyzed by simulation and feasibility tests with the Cheddar tool [4]. The Open Source AADL Tool

---

[1] Quality attributes are also referred to as non-functional properties in the literature.

Environment (OSATE)[2] supports resource allocation analysis, while the Ocarina toolset[3] allows searching for deadlocks and un-initialized variables. As far as quantitative dependability analyses are concerned, to the best of our knowledge, the only tool reported in the literature is a fault tree generator prototype, proprietary of Honeywell [5]. It only targets reliability and safety measures and is not well suited for obtaining other dependability measures, such as availability, if the components' behaviors are not stochastically independent. Our prototype tool ADAPT (*from AADL Architectural models to stochastic Petri nets through model Transformation*), which is presented in this paper, aims at facilitating the evaluation of various dependability measures (such as reliability and availability) from AADL models. It is based on model transformation rules, from AADL to Generalized Stochastic Petri Nets (GSPNs). The use of GSPNs has several advantages: (1) they can be automatically converted into Markov chains that are very powerful in capturing stochastic dependencies between components and in the evaluation of dependability measures, (2) they allow modular and hierarchical modeling for component-based systems, similarly to AADL, and (3) they provide means for structural verification of the model. Such verification support facilities are very useful when dealing with large models. The AADL architectural model given as input to ADAPT can be used unmodified for several complementary analyses, such as those mentioned above, which enables making tradeoffs during system design with respect to various view points.

The remainder of this paper is structured as follows. Section 2 gives a brief overview of AADL. Section 3 is dedicated to the principles that guided the definition of the model transformation rules from AADL to GSPN. Section 4 presents ADAPT from the developer's and from the user's perspective. Section 5 summarizes the paper and presents perspectives for improving ADAPT.

## 2. About AADL

In the AADL, systems are modeled as hierarchical collections of interacting application components (processes, threads, subprograms, data) and a set of execution platform components (processors, memory, buses, devices). The application components are bound to the execution platform. Dynamic aspects of system architectures are captured with the AADL operational mode concept.

The analysis-related information is described separately and then plugged into the architectural model. In particular, the AADL Error Model Annex [6] has been defined and standardized to complement the AADL core language in support to describing dependability-related information (such as faults, fault propagation, repair, fault-tolerance strategies). The AADL architectural model is annotated with error model constructs in order to describe the behavior of components and connections in the presence of faults.

An error model is a stochastic automaton declaring states, events, propagations and transitions between states. Transitions are triggered by events or propagations, which are directional (`in` or `out`). Events and propagations are characterized by Occurrence properties (fixed probabilities or distributions). An outgoing propagation of a component is matched to an incoming propagation of another component if the components are connected or bound at the architectural level. When an error model is associated with a component, it is possible to customize it by setting component-specific values for the occurrence for error events and error propagations declared in the error model. It is also possible to filter propagations by using `Guard` properties.

The behavior of the system is obtained by composing the individual behaviors of components, according to *dependency rules* specified in the AADL Error Model Annex [6]. Such dependencies may result for example from fault propagations between components, or from functional of structural interactions. The tool ADAPT traverses the AADL architectural model to search for error models and uses these dependency rules to establish correspondences between `out` propagations of a particular component and name-matching `in` propagations or `Guard` properties of other components.

## 3. On the model transformation

ADAPT supports the modeling framework published in [7]. This framework is formed of an iterative modeling approach with modeling guidelines, reusable patterns for fault-tolerant architectures and a model transformation from AADL to GSPN that allows obtaining dependability measures from the AADL model. ADAPT implements the set of transformation rules from AADL to GSPN presented and formalized in [8]. The set of model transformation rules has been designed to be automated. Also, the definition of the rules favors the modularity of the GSPN. Indeed, the resulting GSPN is structured as a set of subnets: *component subnets* that model the behavior of components in the presence of their own

---
[2] http://www.aadl.info/OpenSourceAADLToolEnvironment.html
[3] http://ocarina.enst.fr

faults and repair events, and *dependency subnets* that model the dependencies.

We defined two sets of transformation rules. The first set is devoted to the transformation of the AADL models of the components, to create the component subnets: the components' error models are processed by taking into account their states and transitions, triggered by events. The second set of rules is related to the transformation of the dependencies between the system components (i.e., functional, structural, maintenance and fault-tolerance). To this end, we have identified all AADL constructs necessary for describing dependability-related dependencies, and we have defined modeling rules for each type of dependency. We have then defined transformation rules for all these constructs. Thus, the resulting set of rules is necessary and sufficient for obtaining a GSPN describing all the identified types of dependencies.

Dependencies are usually described by name-matching **in** - **out** propagations. In a first step, **out** propagations are identified. Then, for each **out** propagation, the AADL architectural model is traversed in order to find **in** propagations in other components that occur as effects of the **out** propagation. The name-matching **in** - **out** propagations are then transformed to obtain dependency subnets.

The subsequent rules are devoted to transforming propagation filtering and masking mechanisms (i.e., **Guard_In** and **Guard_Out** properties), mechanisms for connecting error states to operational modes (i.e., **Guard_Event** and **Guard_Transition** properties, **activate/deactivate** transitions) and hierarchical models (i.e., **abstract** and **derived** error models).

## 4. Overview of ADAPT

ADAPT interfaces the Open Source AADL Tool Environment (OSATE[4]) on the AADL side and SURF-2 [9] on the GSPN side. OSATE is the most used AADL modeling tool. From a developer's point of view, OSATE provides useful methods for traversing and processing the AADL architectural model. In addition to OSATE, we also base our tool on the set of plug-ins developed at the Carnegie Mellon Software Engineering Institute that allow parsing the Error Model Annex[5] constructs.

### 4.1. A developer's perspective

ADAPT is built in the Java programming language on top of the Eclipse IDE[6] (integrated development environment). This implementation choice is due to the fact that we interfaced our tool with OSATE. Other implementation alternatives are recent metamodel-based transformation languages such as ATL [10], MOLA [11], MTL[7] or GReAT [12]. Model transformation techniques supported by them are compared in [13]. ADAPT consists of 10 kilo lines of code, half of which are automatically generated from an Ecore[8] metamodel using the Eclipse Modeling Framework (EMF) [14]. EMF is a modeling framework and code generation facility for building tools and other applications based on a structured data model. From a metamodel specification described in XMI or Ecore, EMF provides tools and runtime support to produce a set of Java classes for the model. We have used this facility to automatically generate the Java classes for handling GSPN elements (see Section 4.1.1).

Figure 1 presents an overview of ADAPT: its structure and interfaces with AADL and GSPN tools respectively. ADAPT is depicted in dark gray together with its outputs. The AADL and GSPN tools it interfaces are shown in light gray. The black dotted arrows represent interactions of type "depends on", e.g., ADAPT depends on OSATE and the Error Model Annex plug-ins.

ADAPT is structured as a set of three Eclipse plug-ins:
1) gspnModel plug-in: contains methods for the creation and handling of GSPN objects.
2) dependency plug-in: contains methods for identifying the existence of dependencies in the AADL model.
3) aadl2gspn plug-in: implements our transformation rules. This is the main plug-in of ADAPT. It depends on the methods implemented in the dependency plug-in to handle the AADL model and identify possible dependencies. It also depends on the gspnModel plug-in to build the GSPN.

The generated GSPN is saved in two forms: a generic XML/XMI file and a tool-specific file complying with the file format of the dependability evaluation tool SURF-2. Both files are obtained from the same GSPN object model, internal to ADAPT. The tool-specific file may also be obtained directly from the XML/XMI file. Possible interfaces with other GSPN-based dependability evaluation tools are represented with dashed arrows.

---

[4] http://www.aadl.info/OpenSourceAADLToolEnvironment.html
[5] http://www.aadl.info/downloads/errormodel-1.1.6/osate-errormodel-frontend-1.1.6-08142007.zip
[6] http://www.eclipse.org/
[7] http://modelware.inria.fr/article66.html
[8] Ecore is a small and simplified subset of UML, used in the Eclipse Modeling Framework.

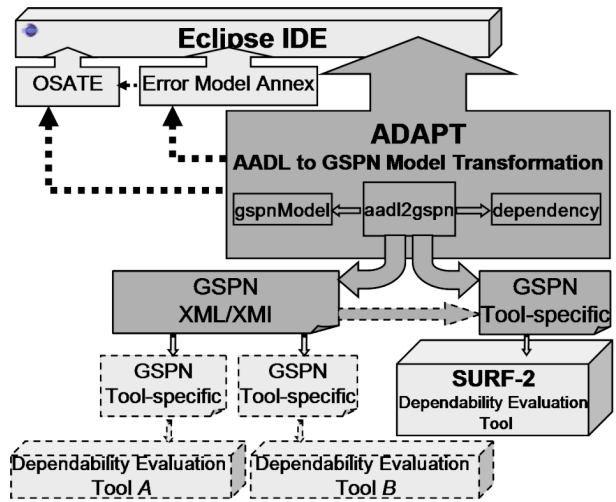

**Figure 1. Overview of ADAPT**

The three plug-ins forming our model transformation tool are described successively in subsections 4.1.1, 4.1.2 and 4.1.3.

**4.1.1. gspnModel plug-in: Ecore metamodel.** This plug-in offers all the methods necessary for creating and customizing GSPN elements (places, transitions and arcs) and for traversing a GSPN model. The code of this plug-in has been automatically generated from an Ecore metamodel of GSPN using the EMF. The GSPN built by ADAPT is compliant with this metamodel. An XML/XMI schema is also generated from the Ecore metamodel. ADAPT saves the GSPN under XML/XMI format, which is compliant with this schema.

Figure 2 shows the Ecore metamodel used by the Eclipse Modeling Framework to generate the code. *PetriNet* object contains several *Arcs* and several *PlaceOrTransition* elements. *Arcs* are described by a weight while *PlaceOrTransition* elements are identified by names. *Arcs* and *PlaceOrTransition* elements cannot be instantiated directly (they are abstract). Concrete arcs of types *PlaceToTransition* and *TransitionToPlace* can be instantiated and inherit from the *Arc* elements. *Place* and *Transition* elements inherit from the *PlaceorTransition* elements. A *Place* is characterized by an initial marking. A *Transition* is characterized by an Occurrence type and a parameter. Associations are established between the *TransitionToPlace* / *PlaceToTransition* arcs and *Place* and *Transition* elements.

**4.1.2. dependency plug-in.** This plug-in is a library of methods aimed at identifying possible dependencies between components of the AADL model. The details of this library are presented in [15]. From a practical point of view, this plug-in implements the elementary dependency rules specified by the AADL Error Model Annex. They determine the possible dependencies between error models associated with components and connections, based on the various interactions in the architectural model. Besides a few special cases, most of the interactions fall into the three following categories:

1) They may be due to the fact that application components run on top of platform components. For example, out propagations declared in an error model associated with a processor are visible in all threads bound to that processor.
2) They may be due to the fact that application components interact through connections, accesses to shared data and calls to services provided by other components. For example, out propagations declared in an error model associated with a component can impact all components reachable through connections.
3) They may be due to the fact that platform components are connected to each other through shared access to buses. For example, out propagations declared in an error model associated with a bus arrive to all components accessing the bus.

The methods implemented in this plug-in allow identifying receiver and sender components or connections for a given error propagation declared in an error model associated with a component or a connection of the system instance. Once the dependencies are identified, the *aadl2gspn* plug-in takes over to perform the model transformation.

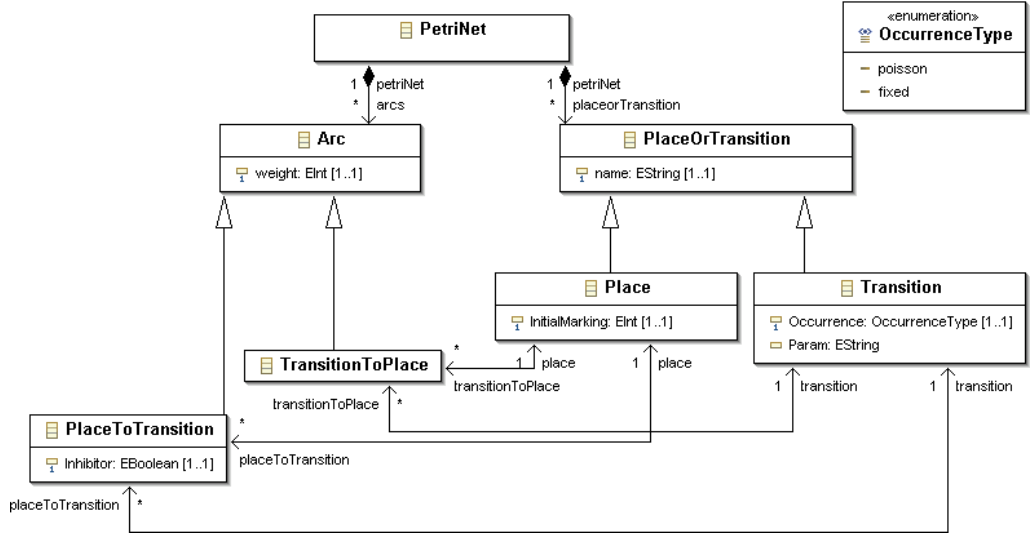

**Figure 2. Ecore metamodel for GSPN**

**4.1.3. aadl2gspn plug-in.** This is the main plug-in of our tool. It performs the transformation of error model elements into GSPN elements, according to the rules presented in [8]. It implements a metamodel-based transformation. It uses the metamodels of the AADL [16] and of the AADL Error Model Annex [17] as a source and of the GSPN as a target.

The transformation is performed iteratively. First, the model transformation rules for independent components are applied. Then possible dependencies are identified and the transformation rules for dependencies are applied.

## 4.2. A user's perspective

An OSATE user installs ADAPT as an Eclipse feature and a set of plug-ins. ADAPT requires that the Error Model Annex support plug-ins, provided by the Carnegie Mellon Software Engineering Institute, be installed too. In order to run the AADL to GSPN transformation tool, the user must instantiate an AADL system model, and select the resulting system instance. The system instance must have an associated `Derived_State_Mapping` expression that we use to derive the state partitions necessary to the dependability evaluation tool, to evaluate measures. A `Derived_State_Mapping` expression represents the behavior of a component in the presence of faults in terms of global states as a *logic expression* of the states of its subcomponents.

The `Derived_State_Mapping` expression associated with the system instance must explicitly define the *Failed* global state of the system instance as a Boolean expression of states of its components. If safety is among the targeted measures, a *Catastrophic* global state must also be defined.

The ultimate goal is to obtain quantitative dependability measures. Thus, ADAPT requires that all events and propagations have Occurrence properties (fixed probabilities or Poisson distributions). If an event or a propagation does not have an Occurrence property specified, ADAPT assumes it is immediate of probability 1.

The GSPN obtained after transformation is saved in two files with different formats:
- a generic XML/XMI file, which is a gateway for interfacing other dependability evaluation tools with a minimum amount of effort.
- a tool-specific file that can be imported in the dependability evaluation tool SURF-2. SURF-2 allows the user to customize the model by giving particular values or value ranges to model parameters corresponding to symbolic Occurrence properties coming from the AADL model. The user is also required to define rewards and the measures of interest.

## 5. Summary and future work

This paper presented ADAPT, a model transformation tool whose input is an AADL architectural model annotated with dependability-related information and whose output is a dependability evaluation model in the form of a GSPN. The tool interfaces OSATE on the AADL side and SURF-2 on the dependability evaluation side. Also, it can be easily interfaced with other GSPN-based dependability evaluation tools as it generates a GSPN in a generic XML/XMI format. It is noteworthy that ADAPT is available upon request as open-source, so that it can be reused for further AADL-related developments.

ADAPT is built as a set of plug-ins on top of the Eclipse platform. In the current prototype, all transformation rules presented in [8] are implemented, except for the rules for `activate`/`deactivate` transitions and `derived` error models. As a consequence, it is assumed that the behavior of the system in the presence of faults is identical in all operational modes. Future work includes the implementation of the remaining rules.

We have used the current ADAPT prototype to transform the AADL dependability model of a subsystem of the French Air Traffic Control System including two hardware components sharing a repairman, a fault-tolerant software unit and eight dependencies of several types.

Finally, it is worth to mention that PNML (Petri Net Markup Language) [18] is intended to become an extensible interchange standard for Petri nets. Our work can be easily extended by using the PNML instead of the rather simple meta model for GSPN illustrated in Figure 2.

**Acknowledgements**. This work is partially supported by 1) the European Commission (ReSIST Network of Excellence No. IST 026764), 2) the European Social Fund and 3) Zonta International Foundation.

## 6. References

[1] SAE-AS5506, "SAE Architecture Analysis and Design Language (AADL)," International Society of Automotive Engineers, Warrendale, PA, USA November 2004.

[2] J.-P. Blanquart, A. Rossignol, and D. Thomas, "Toward Model-Based Engineering for Space Embedded Systems and Software," presented at 3rd European Congress on Embedded Real Time Software, Toulouse, France, 2006.

[3] O. Sokolsky, I. Lee, and D. Clarke, "Schedulability Analysis of AADL Models," presented at 20th Parallel and Distributed Processing Symposium, Rhodes Island, Greece, 2006.

[4] F. Singhoff, J. Legrand, L. Nana, and L. Marcé, "Scheduling and Memory Requirements Analysis with AADL," presented at SIGAda Int. Conf. on Ada, Atlanta, GE, USA, 2005.

[5] A. Joshi, S. Vestal, and P. Binns, "Automatic Generation of Static Fault Trees from AADL Models," presented at Workshop on Architecting Dependable Systems of The 37th Annual IEEE/IFIP Int. Conference on Dependable Systems and Networks, Edinburgh, UK, 2007.

[6] SAE-AS5506/1, "SAE Architecture Analysis and Design Language (AADL) Annex Volume 1, Annex E: Error Model Annex," International Society of Automotive Engineers, Warrendale, PA, USA June 2006.

[7] A. E. Rugina, K. Kanoun, and M. Kaâniche, "A System Dependabiliy Modeling Framework using AADL and GSPNs," in Architecting Dependable Systems IV, vol. 4615, LNCS, R. de Lemos, C. Gacek, and A. Romanovsky, Eds.: Springer-Verlag, 2007, pp. 14-38.

[8] A. E. Rugina, "Dependability Modeling and Evaluation - From AADL to Stochastic Petri Nets," in Systèmes Informatiques. Toulouse: PhD dissertation, Institut National Polytechnique de Toulouse, November 2007, pp. 151.

[9] C. Béounes, M. Aguéra, J. Arlat, S. Bachmann, C. Bourdeau, J.-E. Doucet, K. Kanoun, J.-C. Laprie, S. Metge, J. M. d. Souza, D. Powell, and P. Speisser, "Surf-2: a program for dependability evaluation of complex hardware and software systems," presented at 23rd IEEE Int. Symposium on Fault Tolerant Computing, Toulouse, France, 1993.

[10] F. Jouault and I. Kurtev, "Transforming Models with ATL," presented at Model Transformaion in Practice Workshop at ACM/IEEE International Conference on Model Driven Engineering Languages and Systems Montego Bay, Jamaica, 2005.

[11] A. Kalnins, J. Barzdins, and E. Celms, "Model Transformation Language MOLA," in Model Diven Architecture, vol. 3599/2005, LNCS, U. Asmann, M. Aksit, and A. Rensink, Eds.: Springer, 2005, pp. 62-76.

[12] A. Agrawal, G. Karsai, and F. Shi, "Graph Transformations on Domain-Specific Models," Institute for Software Integrated Systems, Vanderbilt University, Nashville, Technical Report 2003.

[13] K. Czarnecki and S. Helsen, "Classification of Model Transformation Approaches," presented at Workshop on Generative Techniques in the Context of Model-Driven Architecture of ACM Conference on Object-Oriented Programming, Systems, Languages, and Applications, Anaheim, CA, USA, 2003.

[14] F. Budinsky, D. Steinberg, E. Merks, R. Ellersick, and T. Grose, Eclipse Modeling Framework: Addison-Wesley, 2004.

[15] M. Bruffa and A. E. Rugina, "A Library Implementing Propagation Rules defined in the AADL Error Model Annex " LAAS-CNRS, Toulouse 07001, February 2007.

[16] SAE-AS5506/1, "SAE Architecture Analysis and Design Language (AADL) Annex Volume 1, Annex C: AADL Meta-Model and Interchange Formats," International Society of Automotive Engineers, Warrendale, PA, USA June 2006.

[17] P. H. Feiler, "Error Model Meta Model and Plug-in," Pittsburgh, PA, USA May 2007, http://la.sei.cmu.edu/aadl/downloads/errormodel-1.1.6/ErrorModelPlugin52007.pdf.

[18] ISO/IEC, "Software and Systems Engineering - High-level Petri Nets, Part 2: Transfer Format," International Standard 15909-2 WD Version 0.9.0 June 2005.